\documentclass[%
 reprint,
 amsmath,amssymb,
 aps,
]{revtex4-1}

\usepackage{graphicx}
\usepackage{dcolumn}
\usepackage{bm}
\usepackage{xcolor}

\begin{document}


\title{Informational entropy thresholds as a physical mechanism to explain power-law time distributions in sequential decision-making}

\author{Javier Crist\' in, Vi\c{c}enc M\' endez and Daniel Campos}

\date{\today}

\begin{abstract}
While frameworks based on physical grounds (like the Drift-Diffusion Model) have been exhaustively used in psychology and neuroscience to describe perceptual decision-making in humans, analogous approaches for more complex situations like sequential (tree-like) decision making are still absent. For such scenarios, which involve a reflective prospection of future options to reach a decision, we offer a plausible mechanism based on the internal computation of the Shannon's entropy for the different options available to the subjects. When a threshold in the entropy is reached this will trigger the decision, which means that the amount of information that has been gathered through sensory evidence is enough to assess the options accurately. Experimental evidence in favour of this  mechanism is provided by exploring human performances during navigation through a maze on the computer screen monitored with the help of eye-trackers. In particular, our analysis allows us to prove that: (i) prospection is effectively being used by humans during such navigation tasks, and a quantification of the level of prospection used is attainable, (ii) the distribution of decision times during the task exhibits power-law tails, a feature that our entropy-based mechanism is able to explain, in contrast to classical decision-making frameworks.
\end{abstract}

\maketitle


\twocolumngrid
\section{Introduction}
In our daily life, we  constantly find ourselves in situations that imply making decisions: what I am going to eat, which film I will see or if I am on time for the next bus. In all these situations we need to evaluate the different options available as a way to elucidate the best one. While exploring such situations would lie within the field of psychology, in the recent years there has been a growing interdisciplinary interest in decision-making. Determining the neural correlates of decision mechanisms constitute an important subject in cognitive and behavioral neuroscience \cite{neural1,neural2,brain,biases}. Also, the mathematical study of decision strategies and its comparison to the subjects performance represents an important subject in game theory and econophysics \cite{game,game2}. Last but not least, ideas from statistical physics and/or complex systems have also made its way; while most contributions to date focus on decision-making at the level of groups or collectives (see \cite{active1,active2,active3,active4,voter1} for some reviews), tentative works suggesting physical principles that could be involved in individual decisions do also exist \cite{ortega2013,yukalov2014,schwartenbeck2015,roldan2015,favre2016}.

Up until now, large efforts have been put in understanding the dynamics and the characteristics of perceptual decisions, it is, those where sensory information provides direct evidence for choosing between the options available, as in the famous random dot motion task \cite{dotmot1,dotmot2}. As a result, a correspondence between such sensory information and the neuronal responses responsible for the evidence accumulation in the brain are assumed to be identifiable somehow. Alternatively, value-based or preferential decision-making (though the exact definition changes from one specific field to another) involves situations where a deliberative and subjective (up to some level) process is necessary to reach the decision, as for example when a subject is asked to choose between two food items. In such cases neural correlates become obviously more difficult to identify. 

We can still introduce a third class of situations in which an objective answer to the task does exist but such an answer cannot be reached instantaneously from sensory information, because successive coupled decisions are involved. Following some existing literature (see, e.g, \cite{tartaglia2017,zhang2017} and references there in), we will denote these situations as \textit{sequential} decision-making. Since these obviously require a higher cognitive capacity and a more reflective response by the subject in order to process the information, these situations are essentially restricted to humans (or maybe some other higher organisms). They include tasks like playing board games as chess, or solving mazes or tasks presented in some intelligence tests. All these examples involve decisions where a tree-like structure of future possibilities must be ideally built. So that, in the present work we will use the term \textit{prospection} to denote such hypothetical, or mental, simulations of future events requiring high memory and abstraction capacities \cite{gilbert2007,suddendorf2007,pfeiffer2013}. 

For the simpler case of perceptual decision-making, most theoretical frameworks aimed to explain their underlying mechanisms and dynamics lie within the so-called \textit{accumulator} framework. In it, cognitive evidence (described through some effective stochastic process) is gained throughout the time until it reaches a given threshold, which then triggers the decision. The paradigmatic example is the Drift-Diffusion model (DDM) \cite{ddm1}, where the relative evidence in favour of the different options is assumed to follow a Brownian diffusion process (which introduces cognitive fluctuations or noise in the process), with a drift that accounts for the trend towards the correct option. Nowadays, it is widely accepted among psychologists that the success of the DDM is overwhelming \cite{ratcliff,rangel}, though in many cases this requires non-trivial modifications or extensions, as time-dependent thresholds \cite{ddm_boundaries} or dynamic changes on the drift  \cite{fontanesi}. Furthermore, recent works have shown that value-based decisions can be also accommodated within this framework provided that the thresholds are assumed to collapse progressively over time \cite{valuev,roxin}.

On the contrary, stochastic mechanisms able to capture the dynamics during sequential decision-making are scarce \cite{nguyen2019} due to their complexity. Here we will provide experimental evidence that those processes in humans, or at least those of a certain type, are compatible with a stochastic framework in which computational information (through Shannon's entropy) could be implicitly computed by the individual as a way to measure the information gathered. To illustrate this, we study the performance of subjects during a particular navigation task through a maze on the computer screen, combined with eye-tracking data to assess the corresponding behavioral dynamics. We do not introduce any explicit costs for prospecting or analyzing information, as there are no time constraints present in the task. Thus we pose an extreme situation where decisions are mostly driven by optimization of the prospection process, rather than by any speed-accuracy trade-off or any other constraint. This represents an idealized scenario in which the underlying process used by the individuals to reach decisions could be observed without the interference from other factors.


In Section \ref{theoretical} we will present our information-based framework and discuss its main conceptual differences with accumulator models used for perceptual decision-making. In Section \ref{sec:res} we will show our experimental results to describe the performance of the subjects in the navigation task. Comparison of those performances to those shown by virtual (random-walk) algorithms able to prospect information ideally, allows us to infer the level of information that humans really process during the task. This reveals that human performances can only be explained if prospection is actually being used in the task, and we can even quantify that level of prospection. Next, we explore the statistical properties of the response time dynamics observed during the task to provide quantitative evidence that humans performances are compatible with the entropy-based mechanism proposed here. The conclusions from these results are then discussed in Section \ref{conclusions}, and the experimental and numerical methods employed for the analysis are detailed in Section \ref{sec:methods}.

\section{Theoretical framework}\label{theoretical}

A relevant problem in decision-making is to establish a criterion to identify when we have enough information to discriminate between alternative options, e.g. options $A$ and $B$ (in a binary case). This can be accounted for by sequential analysis. Let $\mathbf{x_n} = \{ x_1, x_2, \ldots, x_n \}$ be a set of independent observations that provide some information about the options. We want to use this set to test the hypothesis $H_A$ (corresponding to the option $A$ being valid). Then we need a criterion to determine whether the set $\mathbf{x_n}$ provides either a sufficient level of evidence in favor of (or against) $H_A$, or a larger set is required and then additional information must be gathered. The solution to this problem originally developed by Abraham Wald was the well-known Sequential Probability Ratio Test (SPRT), which can be proved to minimize the size $n$ of the set required to accept or reject the hypothesis with a fixed level of reliability \cite{wald48}. Given the set $\mathbf{x_{n}}$, we can map all its information into the joint probabilities $p_{A,n}$ and $p_{B,n}$ (with $p_{A,n}=1-p_{B,n}$ if the two options are mutually exclusive) that we assign to options $A$ and $B$, respectively. The SPRT criterion establishes then from the corresponding cumulative log-likelihood function
\begin{equation}
   W_n= \ln \left( \frac{p_{A,n}}{p_{B,n}} \right)
   \label{eq:sprt}
\end{equation}
that a decision can be reliably taken as soon as $W_n$ exceeds (or falls below) a given threshold ($W_{th}$).  Consequently, the SPRT criterion establishes that there is a minimum amount of evidence required to decide, and actually the DDM can be seen as a particular continuum implementation of it \cite{bogacz2006,wald48}. 

 \begin{figure*}[]
\centering

			\includegraphics[width=1\linewidth]{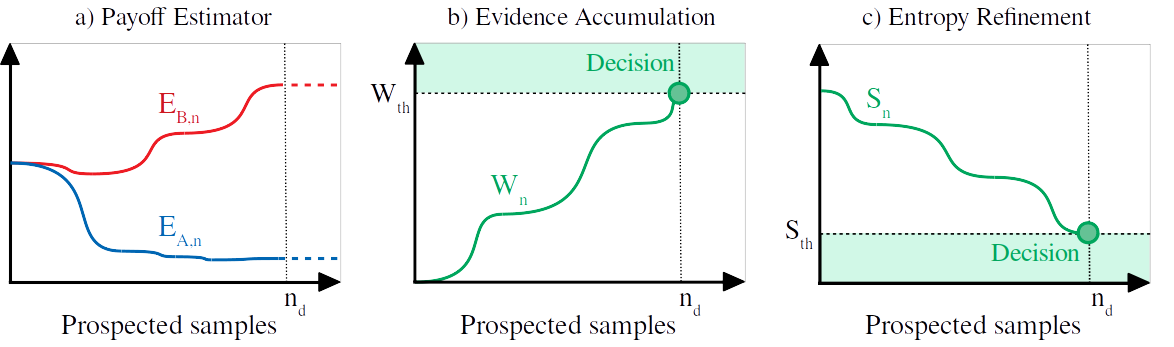}

	\caption{\textit{Scheme for the accumulator and reliability mechanisms. (a) Payoff estimation during successive prospected samples $n$. (b) Wald's ratio $W_n$ evolution according to the payoff estimators (the decision is taken at $n_d$ when $W_n$ reaches the threshold $W_{th}$).  (c) Shannon's entropy $S_n$ evolution according to the payoff estimators (the decision is taken at $n_d$ when $S_n$ reaches the threshold $S_{th}$).}}
	\label{fig:concept}
\end{figure*}

In controlled experiments of perceptual decision-making, the set $\mathbf{x_n}$ corresponds to direct sensory evidence that can be mapped into the probabilities $p_{A,n}, p_{B,n}$ in a relatively easy manner. For example, $\mathbf{x_n}$ can typically account for visual evidence in favour of one of the two options gathered by the subject during the task. However, in sequential decision-making the existence of such a mapping is far less obvious. Still, some relation between sensory evidence and the internal (mental) assessment of probabilities for options $A$ and $B$ is expected to be carried out by the individual. To simplify, we start by assuming that in sequential-decision making subjects mainly use sensory evidence as a way to estimate the average payoff associated to each option. Following, as above, the binary example for simplicity, we denote such estimated payoffs as $E_{A,n}$ and $E_{B,n}$ after the information in $\mathbf{x_n}$ has been gathered. If we are able to find a criterion to determine how the estimations $E_{A,n}$ and $E_{B,n}$ are carried out, this can be translated into a probability map using the Maximum Entropy Principle (MEP) from information theory. According to the prescriptions from the MEP \cite{jaynes1957}, if the only information available we have from a stochastic variable (an estimation of a payoff, in this case) consists of its average $E_{i,n}$ (with $i={A,B}$), then the most neutral (or unbiased) choice of a probability map $p_{i,n}=p_{i,n}(E_{i,n})$ we can build out of it reads
\begin{equation}
    p_{i,n}= \frac{e ^{\beta E_{i,n}}}{Z_n},
    \label{canonical}
\end{equation}
where $\beta$ is a positive constant (which appears as a Lagrange multiplier when applying the formalism of the MEP) and $Z_n$ a normalization factor that guarantees that $\sum_i p_{i,n} =1$ holds.

Note that this formalism is equivalent to canonical or Maxwell-Boltzmann statistics in statistical physics (except for a minus sign in the exponential, that can be absorbed in the definition of $\beta$). Interestingly, combining (\ref{eq:sprt}) and (\ref{canonical}) leads to $W_n= \beta (E_{A,n} - E_{B,n})$, so the SPRT can be interpreted in this context as a criterion that imposes a threshold in the difference between the estimated payoffs to take the decision.

In perceptual decisions for which $\mathbf{x_n}$ translates easily into an estimation of probabilities, and time constraints are strong (these are the most typical experimental conditions used), the criterion to minimize the size of the data-set $\mathbf{x_n}$ given by the SPRT represents an adequate solution. However, when time is not a significant constraint and the decision process requires a slow and reflective processing, as in sequential decision-making, then alternative mechanisms should be explored. 

Here, we argue that a plausible mechanism for such situations must be based on assessing the amount of information that the probability map (\ref{canonical}) contains. The most direct way for computing such information is obviously Shannon's entropy $S_n= - \sum_i p_{i,n} \log{(p_{i,n})}$ (where again $i={A,B}$ for the simplest case of binary decisions). Then, we hypothesize that the easiest way to address sequential decision-making would be to impose a threshold in entropy, $S_{th}$, such that the condition $S_n<S_{th}$ (with $S_n$ obtained, as explained above, from the evidence available) will trigger the decision in favor of the most likely option at that moment. At this point, we remember that Shannon's entropy reaches its maximum value when no information is still available (so $p_{A,n} = p_{B,n}$), and its value decreases as long as higher evidence in favour of one particular option is gained. 

Then, the \textit{evidence accumulation} mechanism typically associated to the SPRT is here replaced by an \textit{entropy refinement} mechanism (see Fig. \ref{fig:concept}). Actually, we note that this idea is not completely novel but other authors have discussed before similar ideas \cite{entromec1,entromec2,entromec3}, though to our knowledge this specific criterion and its implications have never been tested experimentally.

 \subsection{Working example}

\begin{figure*}[]
\centering

			\includegraphics[width=0.99\linewidth]{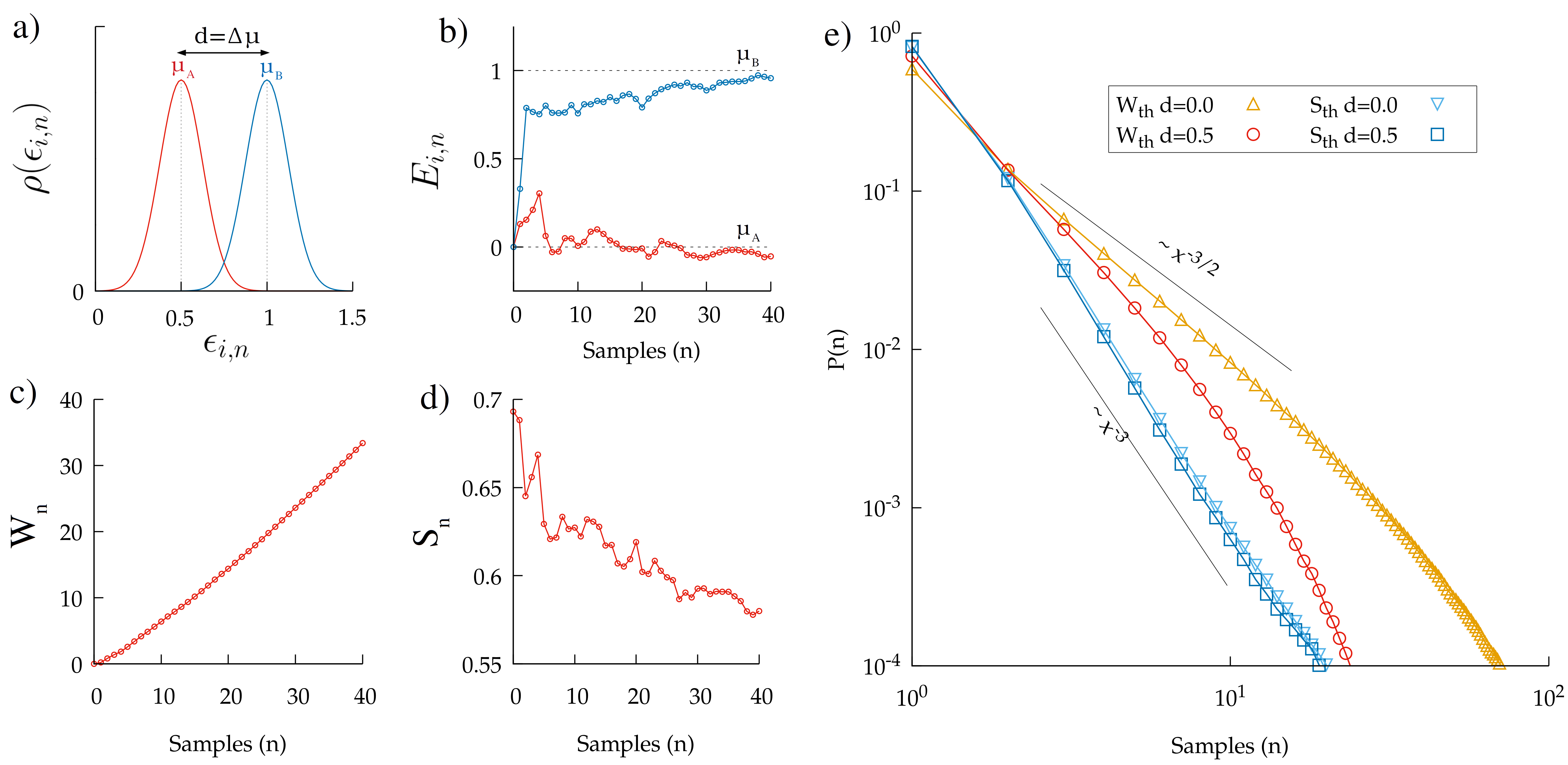}

	\caption{\textit{a) Probability distributions for the stochastic variables $\epsilon_{i,n}$ (with $n$ the number of samples and where $i$ labels the options $A$ and $B$). The means $\mu_{A}$, $\mu_{B}$ represent the actual payoffs for each option. b) Evolution of the estimator $E_{i,n}$  as a function of the number of samples $n$. c) Evolution of the cumulative $W_n$ with the number of samples $n$. d) Evolution of the Shannon's entropy $S_n$  with the number of samples $n$. e) Probability distribution for the number of samples to reach $S_{th}$ or $W_{th}$ for the ERT and the SPRT, respectively, and for different distances $d \equiv \mu_{A} - \mu_{A}$.}}
	
		
	\label{fig:threshold_comparison}
\end{figure*}

We will illustrate some specific properties of the \textit{entropic refinement test} (ERT) through an idealized working example. Our main purpose is to detect some relevant (experimentally measurable) differences in the decision-time dynamics that allow us to discriminate between the ERT and the SPRT.
 
We will focus on the binary decision case again (though we will relax this assumption below). If the individual has to choose between options $A$ and $B$ (whose actual payoffs read $\mu_A$ and $\mu_B$, respectively), this will be done by sampling successively information from the two options to obtain estimates $E_{A,n}$, $E_{B,n}$ of the corresponding payoffs (with $n$ again representing the number of steps, or samples). At the $i$-th step the piece of information obtained by the individual will consist of two Gaussian variables $\epsilon_{A_i}$, $\epsilon_{B,i}$ with corresponding means $\mu_A$ $\mu_B$, respectively, and unit variance. Then, the information obtained provides an approximation to the actual values $\mu_A$, $\mu_B$, and the estimated payoffs can be computed through the average over the information sampled, so $E_{A,n}= \frac{1}{n} \sum_{i=1}^{n} \epsilon_{A,i}$ and $E_{B,n}= \frac{1}{n} \sum_{i=1}^{n} \epsilon_{B,i}$. Note that this is in agreement with our assumption above that the individual essentially uses sampling of information to obtain an averaged estimation of the actual payoffs.

Once we have the estimations for the payoffs, we can compute (with the help of (\ref{canonical})) the Shannon's entropy over the information sampled, and explore the decision dynamics if a particular threshold $S_{th}$ is used to trigger the decision. The main magnitude we will explore is, as usual in the field, the statistics of decision times, this is, the number of samples $n$ that the individual requires to reach the entropy threshold. Most works in decision-making experiments focus on the average values of the decision time, or alternatively decision time histograms are fitted to gamma distributions \cite{gamma1,gamma2}. From that analysis, however, it would be extremely difficult to discriminate between different decision mechanisms like the ERT and the SPRT, since an appropriate tuning of the parameters could easily lead to similar estimates from both.

Instead, here we will focus on the behavior at the tail of the probability distribution of decision times. Previous works based on ideas similar to the ERT have suggested that this mechanism can account for power-law distributions of decision times \cite{medina2014,medina2015}. This represents a significant qualitative difference to other mechanisms (as the SPRT) where such distributions often decay in an exponential way.

So that, we carry out numerical experiments using the rules above and determine the distribution of decision times for the ERT and the SPRT, as a function of the parameters $\mu_A$, $\mu_B$, and the thresholds $W_{th}$, $S_{th}$ (see Fig. \ref{fig:threshold_comparison}).




 In summary, we find that the SPRT exhibits a time distribution that depends strongly on the distance between the means of the payoffs $d \equiv \mu_A - \mu_B$ (Fig. \ref{fig:threshold_comparison} e)), and for most situations it eventually decays exponentially (though transient power-law behaviors with exponent $-1.5$ are also found).  Instead, for the ERT the distribution exhibits a power law behavior $P(n) \propto n^{-3}$ for a wide range of $d$ and $S_{th}$ values. Remarkably, the power-law behavior with the $-3$ exponent persists when considering decisions between more than two options; in the Supplementary Material file we show equivalent results for decisions between 4 possible options.
 
So, at this point we have at least one qualitative difference we can use to discriminate between the SPRT and the ERT.

\section{Results}\label{sec:res}

Sequential decision making requires a mental processing of the acquired information which can be hard to capture through monitorization of the brain activity. However, simple situations in which behavioral information reflects somehow such information processing can help inferring the actual mechanisms behind. With this purpose, we have designed a particular navigation task through a maze on the computer screen. Commercial eye-trackers are used during the task to determine where the subjects are gazing at, and from that we obtain information about the possible future paths that the subjects are prospecting at each moment.

 \begin{figure*}[t]
	\centering
	\includegraphics[width=0.8\linewidth]{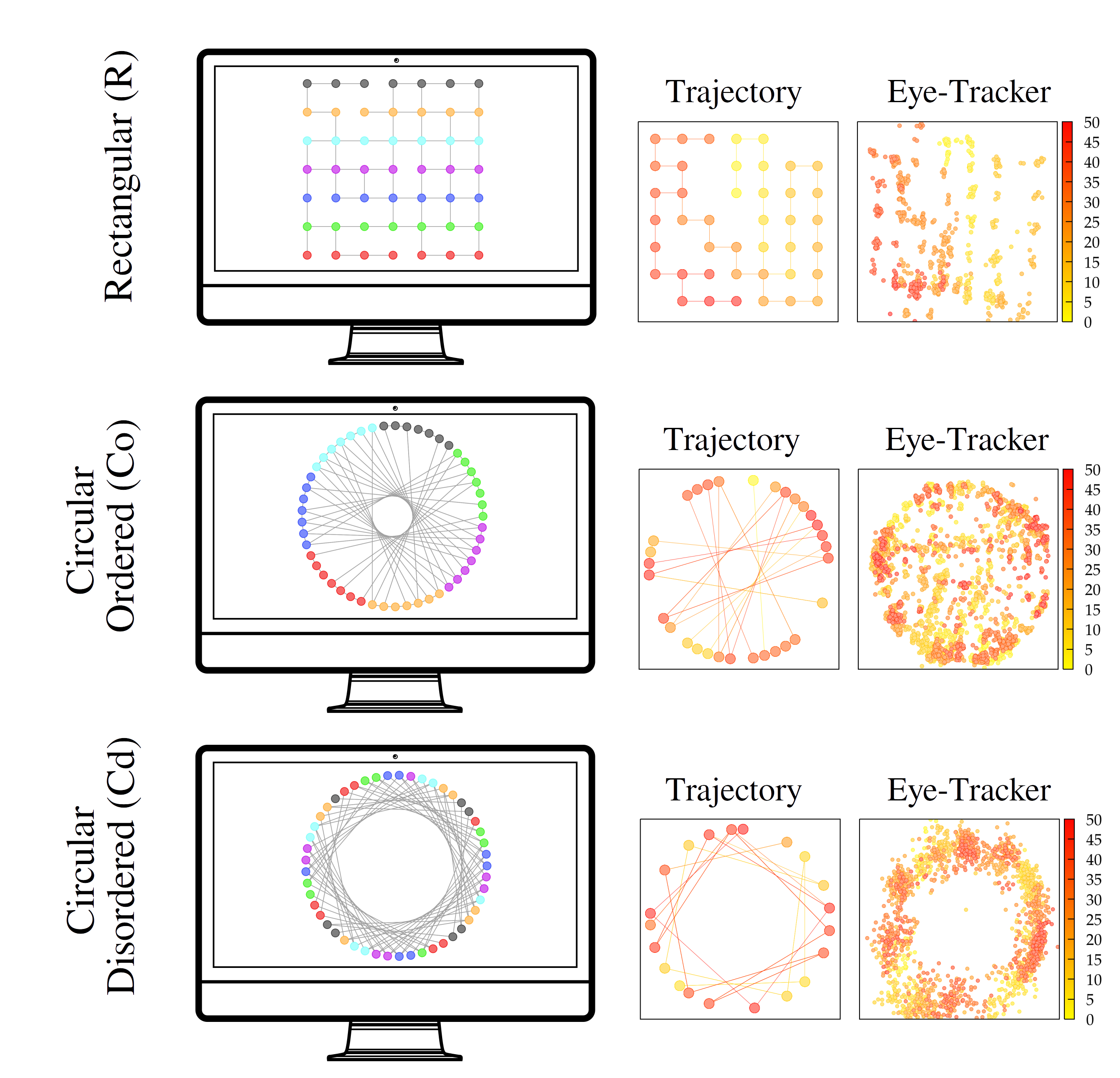}

	\caption{\textit{Scheme of the experimental setup. First column: Visualization of the $49$-node lattice used for the navigation task. The solid lines indicate the bonds allowed between neighbour nodes. Second column: a realization of an individual trajectory within the lattices (the color code denotes time , see legend). Third column: eye fixations obtained during the previous trajectory from eye-tracking data.}}
	\label{fig:triored}
\end{figure*}

The subjects are asked to visit the maximum amount of nodes possible of a discrete 7x7 lattice on the screen within 49 moves if starting from the center of the structure. Moves are only allowed between neighbour nodes, and they are carried out by clicking with the mouse on the node to which one wants to move next. Heterogeneities in the lattice and three different levels of visual complexity (Rectangular, Circular Ordered and Circular Disordered, see left column in Fig. \ref{fig:triored}) are introduced in order to evaluate the subjects performance under different situations. However, all the structures presented to the subjects are topologically identical to facilitate comparison of the results, only visual representation changes from one to another. Further details about the experimental design and protocol are provided below in Section \ref{sec:methods}.


\subsection{\label{sec:perf} Overall performance in the navigation task}

 The overall performance of the individuals is computed as the number of nodes that have been covered during the entire trajectory of 49 moves (Fig. \ref{fig:performance} a)). For the Rectangular level, the subjects visited in average $37.1 \pm 3.8$ nodes (this is, a $75,7 \%$ of the total $49$ nodes). For the Circular Ordered level, they covered $29.1 \pm 4.8$ nodes ($59,4 \%$) and for the Circular Disordered graph, $26.4 \pm 4.8$ nodes ($53,9 \%$). These results confirm that the navigation task (and so the sequential decision making involved) largely depends on the visual representation of the nodes in the lattice, with more complex representations probably preventing the subjects from planning their trajectories ahead (so suppressing prospection). Furthermore, analyzing the performances as a function of the averaged decision time shows us that a higher performance is not a result of spending more time before deciding (Fig. \ref{fig:performance} b)), but the difficulty of the task seems to be clearly the driving force explaining those differences (note that the decision time is here defined as the time between consecutive moves). 



\subsection{\label{sec:time} Eye-tracking data capture prospection dynamics}

\begin{figure*}[]
	\centering
\includegraphics[width=1\linewidth]{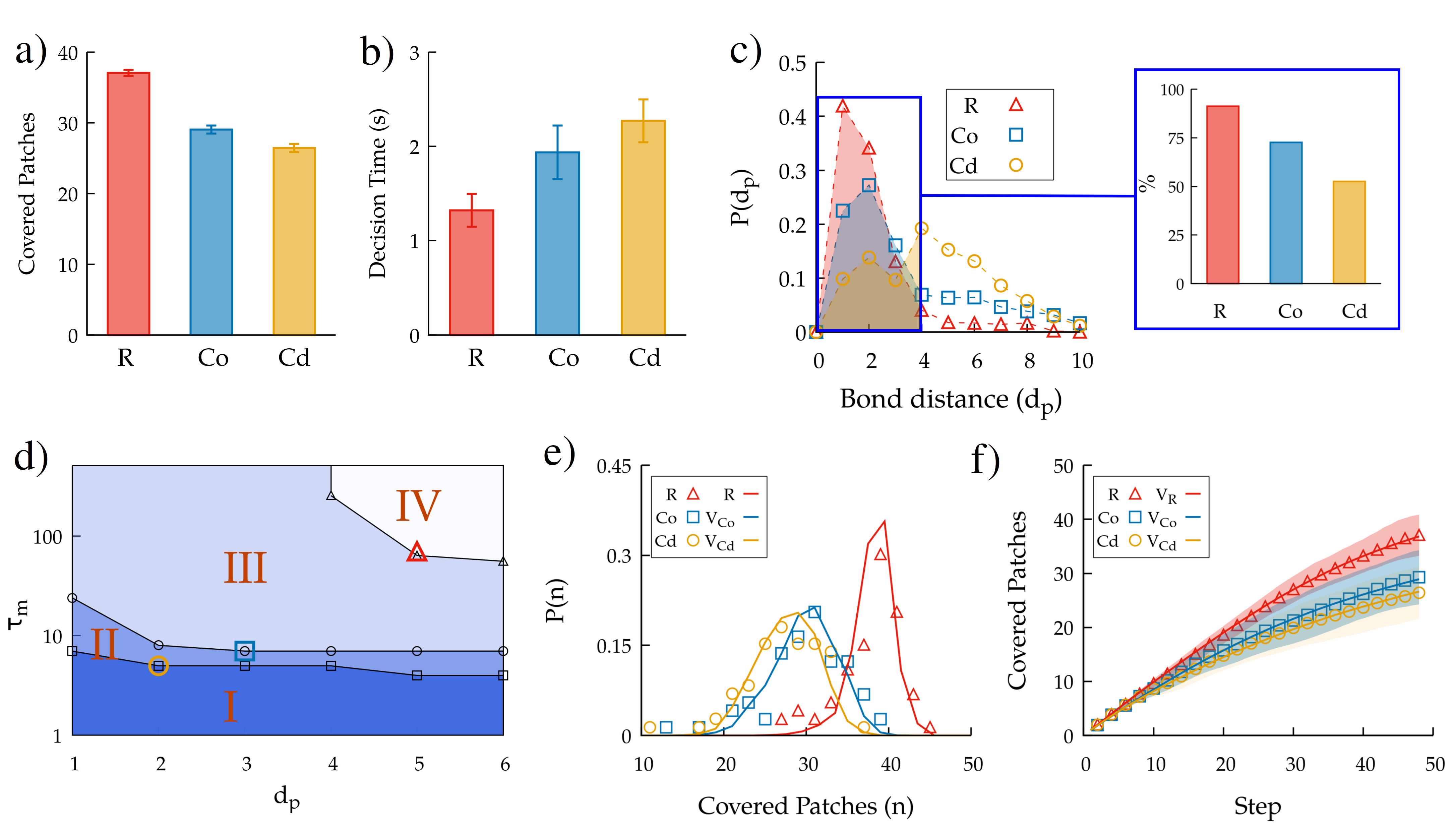}

	\caption{\textit{a) Performance of human subjects in the task for the three levels of visual organization presented in Fig. \ref{fig:triored}. b) Averaged decision times for the three levels. c) Distribution of the distance $d_{b}$ between the current node and nodes gazed between moves (Inset: cumulative probability that the nodes gazed satisfy $d_{b} \leq 4 $). d) Performance of the virtual walker in comparison with experimental ones, with regions II, III and IV accounting for virtual walker performances better than humans in the R, CO and CD cases,respectively. e) Best fit (lines) to the experimental distribution of performances (symbols) obtained from the virtual walker algorithm (see text for details of the fit). f) Evolution of the performance during the $49$-move trajectories obtained from experimental trajectories (symbols) and virtual walkers with best-fit parameters (lines).}}
	\label{fig:performance}
\end{figure*}

We next analyze the information gathering during the task with the help of the eye-tracking data. We define the distance $d_{b}$ as the minimum number of moves required to go from the current node of the lattice to the one the individual is gazing at. The corresponding distributions of $d_{b}$ found are again completely different for the three levels of visual organization (Fig. \ref{fig:performance} c)). Then it is clear that the individuals cannot prospect equally in the three cases. While for the Rectangular level a large amount of time is invested in gazing at nearby nodes, for the two Circular levels (specially for the Disordered one) frequent gazes at distant nodes are observed. These must be attributed either to (i) distractions caused by the presence of nodes which are close on the screen configuration though they are not easily accessible from the current one, or (ii) the difficulty at identifying easily the nodes which are available in the next few steps. Ideally, an efficient prospection of the future paths should combine an intensive exploration of closer nodes and a smaller (but non-negligible) exploration of further ones. We illustrate this in the inset of Fig. \ref{fig:performance} c), where the cumulative probability of gazing at nearby nodes (defined as those with $d_{b} \leq 4$) is shown to decrease drastically as a function of the visual difficulty of the task.

\subsection{Quantifying prospection during navigation}

As a way to quantify and refine the ideas above, we compare the subjects performance in our task to that of virtual subjects following an algorithm which is able to automatically prospect the information of the paths available within a certain number of moves $d_p$ (called the \textit{prospection length}) in the lattice. While a classical random-walk algorithm would select its path completely at random, making uninformed decisions, our virtual subjects (walkers) have the ability to use the information from the paths prospected to avoid revisits to previous nodes as much as possible, using then a mechanism of self-avoidance. Actually, we follow similar rules to those of the true Self-Avoiding Random Walk \cite{tsaw1,tsaw2,tsaw3} and the Self-Attracting Random Walk \cite{asaw1,asaw2,asaw3} schemes to generate the payoffs $E_{j,n}$ (see Section  \ref{sec:methods} for details), so then the procedure described in Section \ref{theoretical} can be applied within the lattice to generate our virtual random walks with prospection. 

To enhance its performance, it is then necessary that the virtual walker keeps in memory its previous trajectory. To implement this in a realistic way, we consider that previous visits to a patch are kept in memory by the walker during a characteristic time $\tau_{m}$. As a result of this finite memory, the corresponding payoffs $E_{i,n}$ will be modified. For large values of $\tau_m$ memory remains untouched, and so all visited sites are remembered. On the contrary, for small $\tau_m$ previous nodes will be forgotten and so all values of $E_{i,n}$ will be always similar, leading to a very homogeneous probability map (note that if all $E_{i,n}$ values are the same, according to (\ref{canonical}) the same probability will apply to all neighbour nodes, and so the virtual walker will behave in a random, uninformed way).

The rules above allow the virtual random walkers to avoid overlaps in their trajectories. However, their performance, contrary to that of human subjects, is independent of the visual organization of the lattice (Rectangular or Circular). Thus, we can use the comparison between both to assess the prospection abilities that are being presumably used by the human subjects in each level of the experiment. 

By exploring a reasonable range of $d_p$ and $\tau_m$ values in the algorithm, we observe that the parameter phase space can be divided into four regions (see Fig. \ref{fig:performance} d)). For region I the algorithm produces an averaged number of visited nodes lower than the individuals in any of the experiments. The region II produces a performance which lies between the results obtained between Circular Ordered and Circular Disordered. The region III overcomes the results for the Circular Ordered performance but not for the Rectangular. The region IV, finally,  outperforms all the experimental results. 

Hence, we conclude that relatively large values of both $\tau_{m}$ and $d_p$ are necessary for the virtual walkers to equal or improve the performance by the subjects in the Rectangular level. This  seems to confirm that the subjects in this case do actually remember the previously visited nodes during the task, and predict future paths efficiently. The prospection ability, in particular, is indispensable to justify the performances seen in the experiments. Instead, for the Circular structures the individuals are probably not able to prospect the paths to distant nodes (information gathering is less efficient, as suggested before in Fig. \ref{fig:performance} c)); in consequence, the value of $d_p$ necessary to reproduce their performance is not necessarily high (though still some level of memory $\tau_m$ is necessary).


Next we determine those values of $d_p$ and $\tau_{m}$ that provide the best fit to the distribution of performances obtained from the experiments (see Fig. \ref{fig:performance} e)). These are (i) $\tau_{m}^{R}=70$, $d_{p}^{R}=5$, (ii) $\tau_{m}^{Co}=7$, $d_p^{Co}=3$, and (iii) $\tau_{m}^{Cd}=5$, $d_p^{Cd}=2$, for the Rectangular (R), the Circular Ordered (Co) and the Circular Disordered (Cd), respectively. 

From this, we analyze the evolution of the performance throughout the task between humans and the virtual walkers with the fitted parameters (Figure \ref{fig:performance} f)). The performances increase almost linearly in the beginning, but the growth slows down as the time advances and trajectory overlaps appear in consequence. The experimental curves (symbols) and those obtained from the virtual walkers (lines) agree almost perfectly. This confirms that  the behavior of virtual walkers with prospection reproduces in detail the dynamic performance of human subjects throughout the experiment, and so it provides a reliable approximation for it.

\begin{figure*}[]
	\centering
	\includegraphics[width=1\linewidth
	]{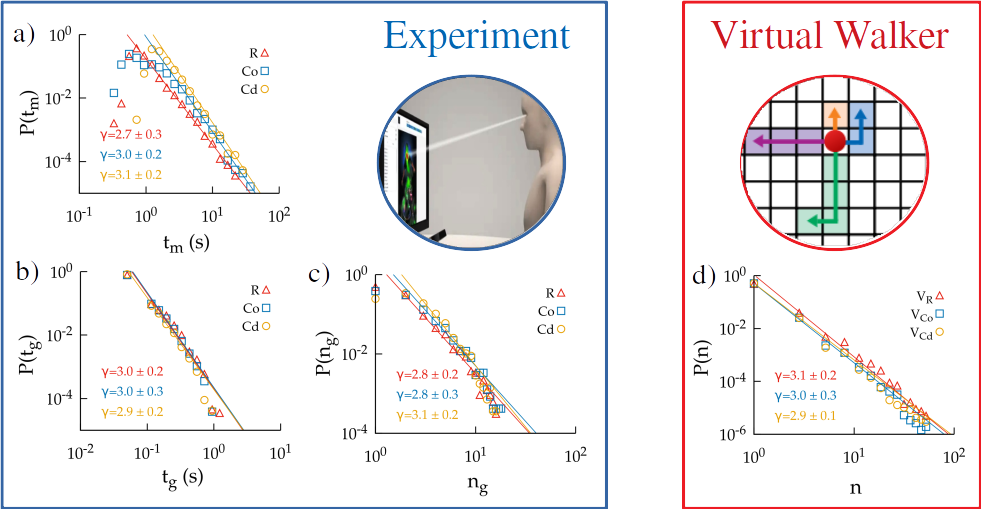}
	
	\caption{\textit{Distributions for time between consecutive moves $t_m$ (a),  the gazing times at a given node $t_g$ (b) and the number of gathered patches between consecutive movements $n_g$ (c) obtained form the experimental data. d) Distribution of the number of prospections $n$ performed by the virtual random walker. In all cases, the exponent obtained from a power-law fit to the distributions is highlighted, with the different colors representing the difficulty levels R, CO and CD.}}
	\label{fig:ojo}
\end{figure*}

\subsection{Human decisions during maze navigation are compatible with the ERT}
\label{dynamics}

The working example explored in Section \ref{theoretical} yields a power-law scaling (with exponent $-3$) for the tail of the decision time distributions within the ERT framework. Actually, this result is not specific of that particular example (based on Gaussian estimations of the actual payoffs). Using the virtual random-walks with prospection described in the previous Section, we obtain exactly the same behavior (Fig. \ref{fig:ojo} d)) under a wide range of parameter values for $d_p$, $\tau_m$ and $S_{th}$, so we can infer that this represents a rather general property of the mechanism proposed (see  Section \ref{sec:methods} for further details).

To check if the performance of human subjects in the navigation task shows also the same scaling, we use now the eye-tracking data from the experiments to analyse the distributions of $(i)$ times between consecutive moves in the experiment, $t_m$, $(ii)$ times during which the subjects gaze at the same patch, $t_g$, and $(iii)$ number of different nodes gazed before making the next move, $n_g$. The first one would represent our best estimation of the decision times in the experiment, while the other two are also provided as alternative measures for the sake of completeness.

The results found show a consistent evidence in favor of a power-law scaling with exponent close to $-3$ for the three cases $t_m$, $t_g$ and $n_g$ (Figs. \ref{fig:ojo} a) to \ref{fig:ojo} c)). Despite the different performances found above (Fig. \ref{fig:performance}) for the subjects in the three levels of organization (Rectangular, Circular Ordered and Disordered), it is remarkable that they all exhibit extremely similar behavior in this case. This suggests that a common underlying mechanism for decision-making is being used by the subjects in the experiment, though the different difficulty of each leads also to differences in the performances. While the range over which the power-law scaling extends is not very wide (since the decision times in the experiment only span through two orders of magnitude) the fits shown are quite robust. Only longer decision times (for which statistics is not very significant since very few decisions extend so much time) show significant departures from it. Furthermore, we noted in Section \ref{theoretical} that the classical SPRT often predicts gamma distributions of decision times with exponential decays, so we remark that this classical framework would be completely unable to explain these results.


\subsection{Information statistics at the moment of the decision}

 \begin{figure*}[]
\centering
\includegraphics[width=1\linewidth]{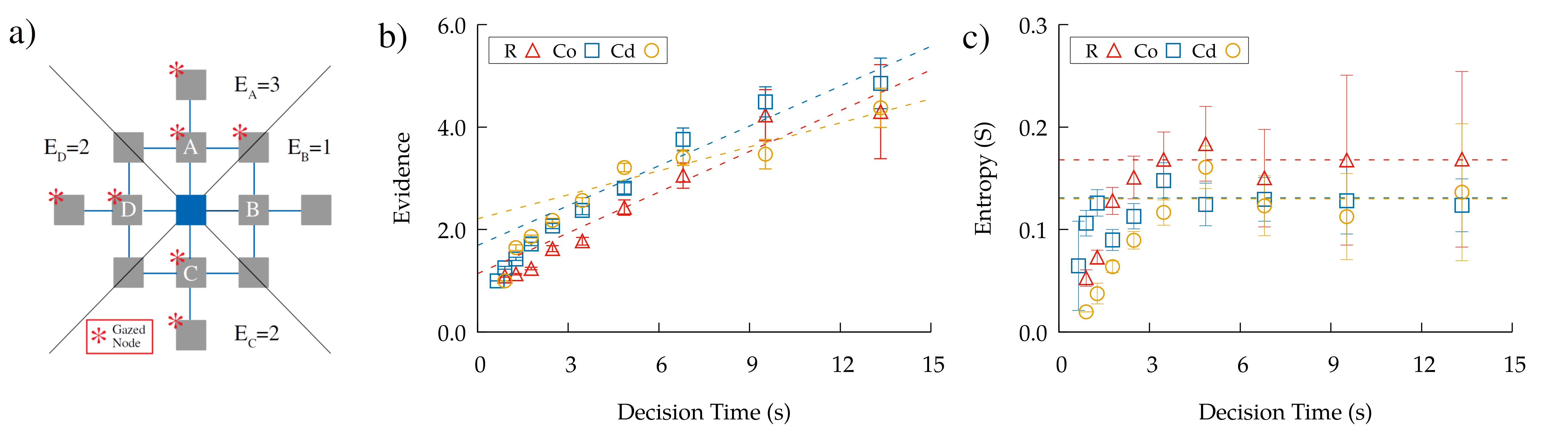}
	\caption{\textit{a) Schematic representation of how to estimate evidence from experimental gazes. The asterisks denote eye fixations so all fixations lying in the same quadrant of one option (e.g. option A) provide evidence in favor of that option. b) Maximum relative evidence between the options at the moment of the decision. Linear fits (for times larger than $3$ s) are given by $f(x)=0.26x +1.14$ (R),$g(x)=0.26x +1.69$ (CO) and $z(x)=0.16x +2.21$ (CD). c) Shannon's entropy $S_n$ at the moment of the decision. The horizontal lines correspond to the averaged entropy for times $>3$ s ($0.168$ (R),$0.131$ (CO) and $0.130$ (CD). Statistical test for the null hypothesis that the entropy is non-constant for times $>3$ s. The corresponding p-values are $p=0.82$ (R), $p=0.69$ (CO) and $p=0.92$ (CD).}}
	\label{fig:entro_ener}
\end{figure*}



To provide further evidence of the compatibility of the experimental results with the ERT against the SPRT, we explicitly plot the estimated cumulative evidence experimentally obtained by the subjects also at the moment of the decision. Since we cannot now which paths the subject is really prospecting, as a proxy for computing this evidence we compute the fraction of time they have been gazing at regions which are in the same direction that the particular node (see Fig. \ref{fig:entro_ener} a)). Starting from the current node, we divide the lattice in four regions, so each eye fixation that lies in a particular region provides further evidence in favour of the corresponding option (A, B, C or D, for the example in Fig. \ref{fig:entro_ener} a)).

As mentioned above, the SPRT criterion with canonical probabilities (\ref{canonical}) is equivalent to assume that the decision is triggered once the relative evidence reaches a given threshold. Our data clearly shows that the relative evidence estimators computed at the moment of making a decision/move increase monotonically with the time that it has been necessary to take the decision. So, longer decisions involve longer evidence accumulation (figure \ref{fig:entro_ener} b)), which is in clear contradiction with the SPRT.

Instead, the ERT proposes that the decision is triggered by a threshold in the informational (Shannon's) entropy $S_n$. When plotting this entropy (computed from the procedure above) at the moment of the decision, it reaches always a value which is approximately constant (independent of the extent of the decision time), suggesting that this magnitude is really an invariant for all the decisions (Fig. \ref{fig:entro_ener} c)). The statistical significance of this result has been verified by testing the null hypothesis that the entropy is non-constant (see figure caption for details). Our conclusion is robust basically for intermediate and longer decisions, while the shorter ones ($<2$ seconds) may be probably induced by an automatic response by the subject, or may be based on prior information gathered during the previous move by the subject, so they step away from the decision dynamics above (actually, the $-3$ power-law scaling discussed above is essentially obtained for decision times in the same range, too).

\section{Conclusions}\label{conclusions}

Navigation efficiency in higher organisms (humans, in particular) must take into account the fact that they are able to prospect the future outcomes of their available options, and process the corresponding information in order to reach a decision. Here we have explored this idea within the context of human navigation through mazes when non-local information is available through visual inspection (and so information is processed in a tree-like fashion).

Our analysis (based on comparing the performances of human subjects and those of virtual walkers with the capacity to prospect future paths) provides evidence that prospection is actually being used by the humans, at least in those levels of visual organization that enable it (the Rectangular one, essentially). Besides, an approximate quantitative characterization of that prospection capacity ($d_p$) and the associated memory skills ($\tau_m$) has been obtained so, reaching an estimation of the quantity of information that humans are really managing during the task.

Furthermore, the distribution of times between moves, or gazing times, together with the study of the values for the entropy at the moment of the decisions allow us to think that the ERT hypothesis proposed here can account to a significant extent for how information is being processed by the subjects during the task. At this respect, we stress that traditionally mean times to decision, as well as the ratio of the times corresponding to choosing option $A$ or $B$ (for binary decisions) have been studied in detail by psychologists. On the contrary, the tails of the decision time distributions are rarely explored decision-making experiments. In contrast, here we have shown that this statistical analysis provides very significant information about the dynamics of decision that is being used in the experiment (and actually non-exponential decays in those distributions clearly seem to indicate that the classical SPRT can hardly be used to explain this kind of tasks/experiments).

Regarding the $-3$ value of the power-law exponent we have recurrently obtained from the ERT formalism and from experiments, a formal justification of its origin remains unknown for the moment. For the specific navigation task proposed here, note that decision times should be understood as the sum of the times that the individual has been gazing at each node before making a new move. Then, to explain the power-law scaling found for decision times one should argue that either (i) the distribution of times the subject keeps looking at a given patch, or (ii) the number of patches that are gazed between decisions, must have power-law tails. It is, however, the case (Fig. \ref{fig:ojo} a) and \ref{fig:ojo} b)) that both distributions present that scaling. So, the underlying mechanism yielding the power-law distribution for decision times is apparently a non-trivial combination of both. It is not still clear yet how general these results may be, or if they appear as a consequence of the conditions used in our experiment in particular. However, we stress that similar results have also been found in other experiments of human navigation through mazes \cite{maze} , so all together raises the need for a deeper and systematic exploration of these ideas in the future.

Finally, it is remarkable that all this information about sequential decision-making in humans has been reached simply with the help of eye-tracking data and monitoring of the decision times exhibited by the subjects on the computer screen, which require just easily available technologies. It is likely that the combination of such methods and data with EEG or other advanced physiological sensors could be used to refine our ideas, and provide more reliable estimates of the dynamics during similar tasks, also in more realistic environments than the one here. We expect that our results can stimulate further research in this line.

\section{\label{sec:methods}Methods} 

 \subsection{\label{sec:exp}Experimental Design}

$18$ clinically normal adults ($10$ women and $8$ men) aged from $18$ to $45$ carried out the experiment. In the first part of the task, subjects are presented a discrete 7x7 regular lattice on the screen (Fig \ref{fig:triored}, upper panel on the left). The patches are linked through bonds connecting them only to neighbour patches (4 paths per node, except for the boundaries where paths are only 2 or 3). However, we remove a part of the bonds between nodes ($20 \%$ of them, always preventing isolated regions in the structure from being formed) in order to introduce some level of heterogeneity in the lattice (Fig \ref{fig:triored}, left column), and then nodes are reorganized in different configurations (Rectangular, Circular Ordered, and Circular Disordered, as mentioned above).

The subjects are asked to visit the maximum amount of patches of the resulting lattice within 49 moves if starting from the center of the structure (one step is defined as a transition between connected nodes in the graph). They are not required to complete the trajectory in any given time, so time constraints are not present in the task and information processing can be extended as much as desired by the subject. They can move to neighbour nodes in the lattice by clicking with the mouse over the patch to which they want to move next (Fig \ref{fig:triored}, middle columns, show some realizations of the resulting trajectories). Heterogeneity in the lattice then makes the process non-trivial (for a homogeneous regular lattice the optimal strategy would be simply to perform a ladder-like trajectory until covering all nodes).

To facilitate visualization of the options available at each decision (especially in the Circular Disordered case, where visualization could be tough), the current node of the individual was depicted in a different color (green, with the rest of the nodes appearing in blue) and the possible moves available at each moment were emphasized (with thicker solid lines). On the contrary, the subjects have no visual guides to distinguish between previously visited and non-visited patches, so they can only use their memory skills to avoid overlaps and increase their performance.

 

To assess the subjects performance under different levels of difficulty, the nodes in the Rectangular lattice are then visually reorganized in a circular way. In one case (Circular Ordered), we keep in the circle the order of the rows of the first rectangular graph (Fig. \ref{fig:triored} middle row). For the other (Circular Disordered), we place the nodes following a circular structure but with random reorganization of nodes (Fig. \ref{fig:triored} lower row). We remark that topologically the three structures are completely equivalent, while visually different. Additionally, we rotated 90º, 180º and 270º the rectangular structure (with their corresponding Circular Ordered and Circular Disordered reorganizations) for randomizing the task (so 12 cases in total, all with the same topological structure, are presented to each subject). The final data-set comprised then $216$ trajectories with a mean duration of $77.1 \pm 2.9$ s each.

As a proxy for information prospection during the task, we use eye fixations measured through a commercial eye-tracker (Tobii X2-30, at 30 Hz).  An eye fixation corresponds to the visual gaze on a single location within the screen. the See the right column in Figure \ref{fig:triored} for a visual trajectory example for each structure. We use this to analyze (i) the number of nodes at which the subject gazes between consecutive steps, and (ii) the time it remains gazing at particular patches. For this, each node was assumed to be represented by a circle of radius $0.05$ (the screen size is equal to $1$) around the center of the node, so all eye fixations lying within the circle are assumed to indicate that the subject is gazing at that particular node. This circle size prevents to assign the subject fixations to different nodes at the same time.

\subsection{Virtual walkers with prospection}

An algorithm for generating virtual random walks with prospection over the 7x7 lattice used in the experiment is proposed as a reference model against which to compare experimental data. Our virtual walkers are able to estimate the convenience of moving to a neighbour node $j$ by assigning successive values $ \epsilon_{j,1}, \epsilon_{j,2}, \ldots$ to that node by prospecting hypothetical paths that would use that node as a starting point. So that, at each time step the walker prospects one particular path (chosen at random from all the possible ones) of fixed length $d_p$ (\textit{prospection length}) starting from each of the neighbour nodes. The specific value $\epsilon_{j,n}$ assigned to the $n$-th prospected path for the neighbour node $j$ corresponds to the fraction of non-visited nodes that the path will cover, with $\epsilon_{j,n}=1$ representing a prospected path for which all sites are still unvisited, and $\epsilon_{j,n}=0$ representing a path for which all nodes have been already visited before. So that, the corresponding payoff associated to that neighbour node $j$ (after $n$ paths have been prospected) reads $E_{j,n} = \frac{1}{n} \sum_{i=1}^{n} \epsilon_{j,i}$, in analogy with the working example discussed above. 

Once payoffs have been defined, the procedure described in Section \ref{theoretical} can then be applied within the lattice to generate our virtual random walks. After each single prospection of one path in each direction, the walker computes the corresponding Shannon's entropy $S_n=\sum_{i}^{n} -p_{j,i} \ln p_{j,i}$; if the computed value falls below a fixed threshold $S_{th}$, the walker makes the decision (it is, a move) according to the probabilities $p_{j,i}$ computed (we have checked that choosing instead the node with the highest probability leads to very similar results). On the contrary, if $S_n >S_{th}$ then the prospection process continues. However, at practice we introduce a rule such that the maximum number of prospections is limited to $100$ to avoid (extremely unusual) situations in which $S_n$ would never decay below $S_{th}$ because all options persistently exhibit very similar payoffs (this rule doesn't significantly modify any of the results reported here).

\textbf{Distributed prospection lengths.}
Assigning a constant prospection length  $d_p$ to all the prospected paths may seem rather unrealistic. Human subjects are expected instead to prospect paths with different lengths depending on the specific situation (complexity, number of choices available, etc). The results in Fig. \ref{fig:ojo} b) also support this, as the number of gazed patches exhibits a variation which spans almost one order of magnitude. 

We have studied then our virtual random-walk algorithm for the case when a distribution of $d_p$ is introduced instead of a constant value. We have tried in particular a distribution $P(d_{p}) \propto \frac{1}{d_{p}^{\gamma}}$ (for $d_p \geq 1$ and with $\gamma$ going from $0$ to $\infty$), with $\sum_{d_{p}=1}^{\infty} P(d_{p}) =1$ to guarantee normalization. The results, which are summarized in the Supplementary Material File, clearly show that the conclusions one obtains so are qualitatively the same as those presented for fixed $d_p$ values in the main text.

\textbf{Robustness of the distribution of decision times on the entropy threshold $S_{th}$.}
We have reported above that the  decision time for the walker exhibits the power-law distribution with exponent $-3$. An analysis to check that this exponent remains approximately constant, independently of the memory and prospection parameters $d_p$ and $\tau_m$, as well as the threshold $S_{th}$, has been carried out using our virtual random-walk algorithm. According to the results found (see the Supplementary Material file), the conclusions reached in the article remain quite robust. Only when very large or very small values of $S_{th}$ are considered (which would represent the case in which decisions are either taken almost immediately without barely any information gathering, or an extremely large amount of information would be necessary to trigger the decision) the $\sim n^{-3}$ scaling breaks down.

\section{Acknowledgements}
This research was supported by the Spanish government
through Grant No. CGL2016-78156-C2-2-R.


%
%
%


\end{document}



\title{Supplementary Material: Informational entropy thresholds as a physical mechanism to explain power-law time distributions in sequential decision-making}

\author{Javier Crist\' in, Vi\c{c}enc M\' endez and Daniel Campos}

\date{\today}

\begin{abstract}

\end{abstract}

\maketitle


\onecolumngrid

\section{Generalization of the toy-model}

In the Main Text we have examined the output of the ERT in an idealized working example. It consisted in a binary decision with options labeled as $A$ and $B$. Here, we extend this idealized working example to a multiple decision situation with options $A$, $B$, $C$ and $D$ (whose actual payoffs are $\mu_A$, $\mu_B$, $\mu_C$ and $\mu_D$, respectively). For this, the individual can sample successively data from the options, and from this data it can obtain estimates $E_{A,n}$, $E_{B,n}$, $E_{C,n}$ and $E_{D,n}$ of the payoffs, following the same procedure as described in the main text. To simplify, we assume that the process to generate the estimations $E_{A,n}$, $E_{B,n}$, $E_{C,n}$ and $E_{D,n}$  is such that at the $i$-th step, or sample, the piece of information obtained by the individual consists of four Gaussian variables $\epsilon_{A_i}$, $\epsilon_{B,i}$, $\epsilon_{C,i}$, $\epsilon_{D,i}$ with corresponding means $\mu_A$, $\mu_B$, $\mu_C$, $\mu_D$ respectively, and unit variance. Then, the information obtained provides an approximation to the actual values $\mu_A$, $\mu_B$, $\mu_C$, $\mu_D$ and the estimated payoff can be computed through the average over the information sampled to date, so $E_{j,n}= \frac{1}{n} \sum_{i=1}^{n} \epsilon_{j,i}$, with $j={A,B,C,D}$. 

Once the estimated payoffs are available, we can successively compute (with the help of equation $2$ in the Main Text) the Shannon's entropy over the information sampled, and explore the decision dynamics if a threshold $S_{th}$ is used to trigger the decision. We explore the statistics of decision times, this is, the number of samples $n$ that the individual requires to reach the entropy threshold. 

We carry out numerical experiments using the rules above and determine the distribution of decision times one typically finds when the ERT are used in this $4$ options scenario. The results in figure \ref{fig:gau_4} confirm that the multi-optional case also reports the same exponent $-3$ for multiple distances between the Gaussians means $d$ that was reported for the binary case in the Main Text. 

 \begin{figure}[h]
	\centering
\includegraphics[width=0.99\linewidth]{figura_gaussianas_4opt_v2.png}

	\caption{\textit{ a) Probability distributions for the stochastic variables $\epsilon_{j,i}$, with $j={A,B,C,D}$. The means $\mu_{A,B,C,D}$ correspond to the real value of each option $A$, $B$, $C$ and $D$. b) Probability distribution for the number of of necessary samples ($n$) to reach the corresponding entropic threshold for different different distances between the means $d$.}}
	\label{fig:gau_4}
\end{figure}

\clearpage

\clearpage

\section{Prospective algorithm} 

\subsection{Definition}

\begin{figure}[h!]
	\centering
\includegraphics[width=0.6\linewidth]{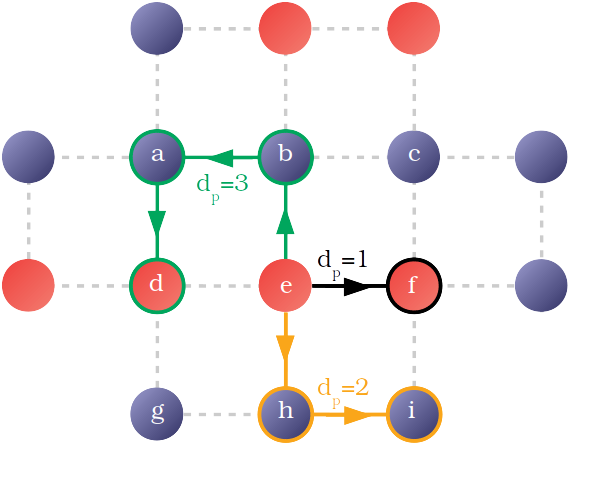}

	\caption{\textit{Scheme of three different prospection paths corresponding to three different prospection lengths $d_{p}=1$ (black), $d_{p}=2$ (yellow) and $d_{p}=3$ (green). The patches that are already visited are marked in red (so they add $0$ to the payoff), while the unvisited ones appear in blue (so they add $1$ to the payoff). The payoff assigned in each one of the paths depicted would be a) $E_{f} = 0$ to option $f$ for $d_p=1$, b) $E_{i} = (1+1)/2 =1$ to option $h$ for $d_{p}=2$, and $E = (1+1+0)/3 =2/3$ to option $b$ for $d_p=3$. }}
	\label{fig:prosp}
\end{figure}

We propose an algorithm in which virtual subjects are able to prospect the paths available within $d_p$ steps in the lattice (we call this parameter the prospection length). For each path prospected, the walker assigns a payoff $E_{i}$ to the neighbour node at which that path starts (for a simple visualization, see Fig. \ref{fig:prosp}). The payoff is taken to be equal to the fraction of visited nodes that the prospected path crosses (so $E$ is bounded between $0$ and $1$, with $E=1$ for a path that does not cover any visited patches, and $E=0$ if all patches covered by the path have been previously visited). 

The walker keeps in memory its previous trajectory during a characteristic number of steps. In particular, the visits are remembered by the virtual subject during a time $\tau$ obtained from the exponential distribution  $P(\tau)=\frac{1}{\tau_{m}}e^{-\frac{t}{\tau_{m}}}$, with $\tau_m$ then representing the characteristic timescale of memory. After this time, the walker will forget that this particular patch has been previously visited and will contribute as a non-visited patch for computing the corresponding payoffs. 

As a result, the algorithm will assign lower payoffs to the options that it remembers having visited and/or that are adjacent to regions that it remembers having visited. So, according to equation $2$ in the Main Text, the probability to choose those options will decrease, so leading the virtual subject to regions that are still unvisited (or at least it does not remember having visited before). A larger prospection length $d_p$ allows the walker to sample the state of further regions and to compute the payoff the information of distant patches, but this will be only efficient if the memory parameter $\tau_m$ is large enough.

Successive prospections of the paths available in each direction are carried out at random among all possible ones of length $d_p$, and so values of the payoffs and the probabilities $p_i$ are continuously updated. Note that for a given value of $d_p$ the number of paths that can be prospected is of the order of $\sim 4^{d_p}$ (if assuming that all bonds between neighbours are available). The algorithm makes the virtual walker to move to one of the available nodes according to the decision criterion described in the Main Text. After each single prospection of one path in each direction, the Shannon's entropy $S=\sum_{i} -p_{i} \ln p_{i}$ is computed; if the value falls below a fixed threshold $S_{th}$, the walker makes a move according to the probabilities $p_i$ computed at that time (we have checked to decide according to the highest probability does not change qualitatively the walker dynamics). On the contrary, if $S>S_{th}$ then the prospection process continues. However, we additionally introduce a rule such that the maximum number of prospections is limited to $100$ to avoid (extremely unusual) situations in which $S$ would never decay below $S_{th}$ because all options available persistently exhibit very similar payoffs. We have carefully checked that this rule doesn't modify any of the results reported in a significant way.

\subsection{Coverage time study}

The results of the algorithm shown in the Main Text have been obtained under the same conditions that in the task presented to the human subjects; this is, for $49$-step trajectories through the $7x7$ lattice with the same topological structure as presented in Fig. 3 in the Main Text. However, for the sake of completeness we also analyze here the dynamics of the prospective algorithm when removing the limitation of $49$ moves, and measuring instead the number of moves it takes to cover all the patches. This gives us an additional insight about the navigation efficiency of the algorithm as a function of the memory and prospection parameters, $\tau_m$ and $d_p$. In particular, we study the mean coverage time ($T_{Cov}$) (this is, the mean time required to cover all sites in the lattice). Minimization of this magnitude would then give an estimation of the navigation efficiency of the algorithm. 

\begin{figure}[h!]
	\centering
	\includegraphics[width=0.69\linewidth]{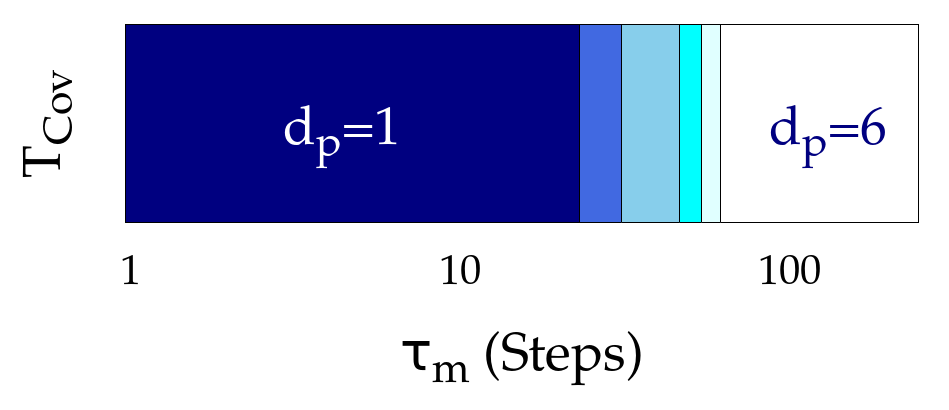}

	\caption{\textit{Prospection length $d_p$ that minimizes the coverage process, i. e., the parameter that provides a performance that minimizes then mean coverage time ($T_{Cov}$) for different values of the memory time $\tau_{m}$.}}
	\label{fig:diag}
	\end{figure}

The main conclusion we can extract (as one can deduce from the results in Fig. \ref{fig:diag}) is that the ability to prospect future paths (so, having a large $d_p$) is useless unless the individual has good memory skills (this is, a large $\tau_m$ value in our context). This makes clear sense, as when the walker cannot remember the previously visited patches (low values of $\tau_{m}$), the optimal strategy consists of removing prospection ($d_{p}=1$); in that case the information provided by further patches represents just useless noise as the walker always sees them as non-visited patches. On the other side, for large $\tau_m$ the walker can correctly identify the previously visited patches (large values of $\tau_{m}$), so then progressively higher prospection lengths $d_{p}$ are found to optimize the coverage of the structure and the search of a target.

\subsection{Distributed prospection lengths}

Assigning a constant prospection length  $d_p$ to all the prospected paths may seem rather unrealistic. Human individuals are expected instead to prospect paths with different lengths depending on the specific situation (complexity, number of choices available, etc). The results reported in fig 5 b) in the Main Text also support this statement (the number of gazed patches is not fixed to a constant number but exhibits a variation which spans almost one order of magnitude). 

We have studied then our algorithm for the case when a distribution of $d_p$ is introduced instead of a constant value. We have tried in particular a distribution $P(d_{p}) \propto \frac{1}{d_{p}^{\gamma}}$ (for $d_p \geq 1$), with $\sum_{d_{p}=1}^{\infty} P(d_{p}) =1$ to guarantee normalization. For $\gamma \to \infty$, the paths are then fixed to $d_{p}=1$, so the prospection algorithm is only to identify whether the neighbour nodes have been visited or not. On the other side, for $\gamma \rightarrow 0$ the probability is uniformly distributed among all $d_p$ values (at practice we limit $d_p$ to $1 \leq d_{p \leq 6}$ since much larger values would be absurd, given the 7x7 maze we have used).  Figure \ref{fig:gamma} reports that sampling a small (but not negligible) number of long paths combined with a majority of short paths (as happens for intermediate $\gamma$ values) is sufficient to recover the same dynamics as obtained for a large fixed  $d_p$ value. This can be seen by comparing results obtained for lower values of $\gamma$ to those of large values of $d_p$, which are extremely similar. This result is remarkable from an evolutionary perspective, since it suggests that improving navigation efficiency would not necessarily require to process much more information continually (note that the number of paths available for prospection grows in general as $n^{d_p}$ for a sequential decision task in which $n$ choices are given to the subject at any step, so processing costs grow exponentially with $d_p$). Instead, having the ability to carry out longer prospections and use this ability just promptly would be enough to increase efficiency significantly.

\begin{figure}[h]
\centering
	\includegraphics[width=1\linewidth]{scores_taus_horizontal.png}

	\caption{\textit{Averaged total number of covered patches after the 49 steps trajectory as a function of the memory time $\tau_{m}$ for the Virtual Walker. The graph a) corresponds to a walker prospecting with a fixed for prospection length $d_p$ and the graph b), to a walker prospecting with a variable prospection $d_p$ obtained from a power-law distribution with exponent $\gamma$.}}
	\label{fig:gamma}
\end{figure}

\begin{figure}[h!]
\centering
	\includegraphics[width=1\linewidth]{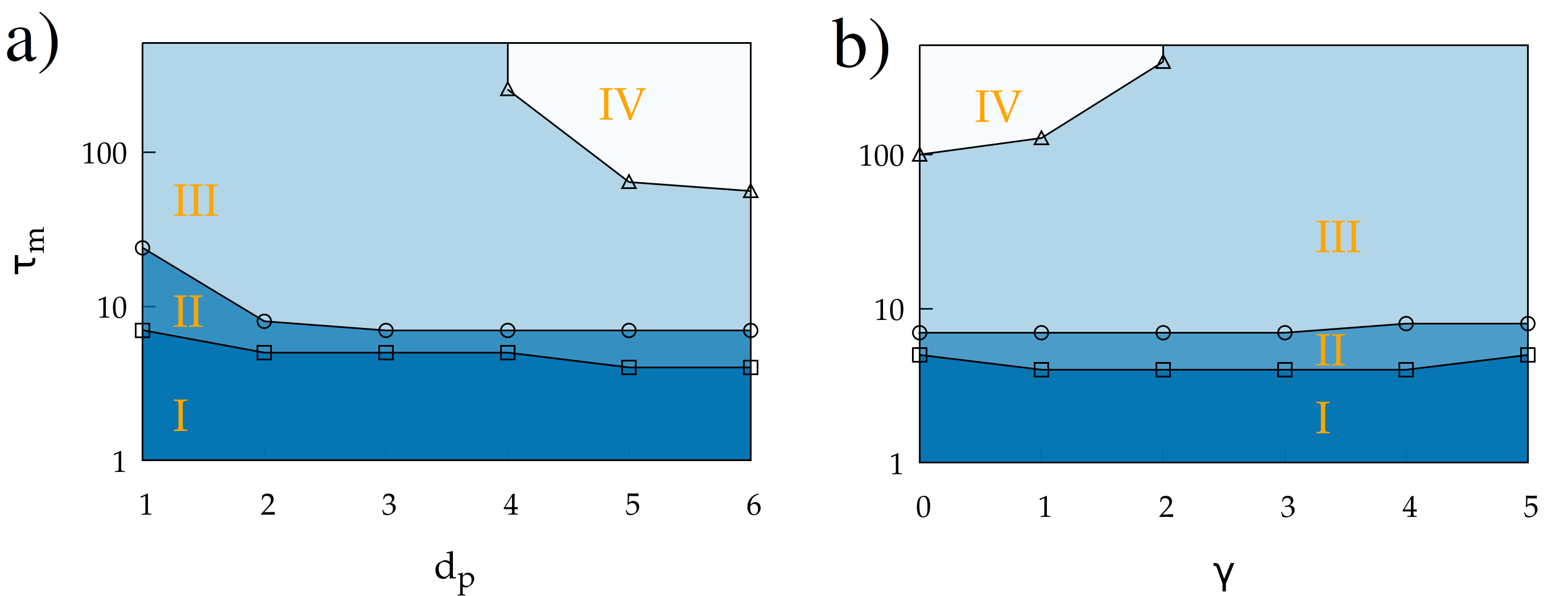}

	\caption{\textit{Diagram for the walker covered patches in comparison with experimental results. Regime $I$ corresponds to a worse averaged performance than all geometries. Regime $II$ corresponds to a better performance than in Circular Disordered.  Regime $III$ corresponds to a better performance than in Circular Ordered and Disordered.  Regime $IV$ corresponds to a better performance than in all geometries. The graph a) corresponds to a walker prospecting with a fixed for prospection length $d_p$ and the graph b), to a walker prospecting with a variable prospection $d_p$ obtained from a power-law distribution with exponent $\gamma$.}}
	\label{fig:comp_diag}
\end{figure}

By exploring the whole range of $\gamma$ and $\tau_m$ values, we can divide the parameter phase space into four regions (figure \ref{fig:comp_diag} b)), analogously as shown in the Main Text for fixed $d_p$. The region I produces an averaged performance that visits less patches than the individuals in any of the experimental graphs. The region II produces a performance which lies between the results obtained between Circular Ordered and Disordered. The region III overcomes the results for the Circular Ordered performance but not for the Rectangular. The region IV, finally,  outperforms all the experimental results. The regions are equivalent to the obtained for fixed path lengths. Again, this shows that distributed values of $d_p$ can be used to obtain higher navigation efficiencies without consuming much higher times of information processing.

\subsection{Robustness of the power-law exponent for the distribution of decision times}

We have reported in the Main Text that the  decision time for the walker, defined as the number of required prospected paths that makes $S < S_{th}$, exhibits again the same power-law distribution (with exponent $-3$) as the Gaussian working example. The results in Fig. 5 in the Main Text correspond to the values $t_{m}$ and $d_p$ obtained from fits to the experimental data. Here we provide an analysis to check that the $-3$ exponent remains as a characteristic feature of the algorithm, independent of the memory and prospection parameters, as well as the threshold $S_{th}$ used in the algorithm. 

First, in Fig. \ref{fig:entropies}a we show the explicit dependence on the entropy threshold, and verify that the power-law behavior is kept as long as reasonable values of this parameter are chosen (extreme choices, with, $S_{th} \rightarrow 0$ for example, would modify the results, but we stress that this represents a rather unrealistic case for the purposes here). While the behavior of the walker is equivalent for different $S_{th}$, we fix it in our results in the Main Text to $S_{th}=0.5$ so it can be conveniently applied to all choices in our maze, regardless the algorithm has two, three or four options available at that movement. On the other side, we observe at Fig. \ref{fig:entropies}b that neither variations in $d_p$ nor $\tau_{m}$ modify significantly the $\sim n^{-3}$ behavior as long as some significant level of memory and prospection is kept. 

We stress that the classical SPRT criterion, as well as other variations we have numerically explored, are unable to reproduce the $-3$ exponent and would lead to much smaller exponents and/or faster (exponential-like) decays in $P(n)$. This, together with the robustness analysis reported here, provides significant robustness to the entropy threshold criterion proposed here.

 \begin{figure}[h!]
	\centering
	\includegraphics[width=1\linewidth]{entropy_threshold_both.png}

	\caption{\textit{a) Distribution of the number of prospections $n$ performed by the walker to force the entropy $S$ to fall below the threshold $S_{th}$.  The parameters $d_p$ and $\tau_m$ are fixed to $d_{p}=6$ and $\tau_{m}=100$, respectively, while different $S_{th}$ are explored. b) Distribution of the number of prospections $n$ performed by the walker to force the entropy $S$ to fall below the threshold $S_{th}$.  The parameter  $S_{th}$ is fixed to $S_{th}=0.5$, while different $d_p$ and $\tau_m$ are explored.}}
	\label{fig:entropies}
\end{figure}


%
%
%
